\documentclass[twocolumn,twoside,slac_two]{revtex4}
\usepackage{graphicx}
\usepackage{fancyhdr}
\begin{document}

\title{Subphotospheric heating in GRBs: analysis and modeling of GRB090902B as observed by Fermi}

%

\author{T. Nymark, M. Axelsson, C. Lundman, E. Moretti, F. Ryde}
\affiliation{Department of physics, KTH, Stockholm, Sweden}
\author{A. Pe'er}
\affiliation{Harvard-Smithsonian Center for Astrophysics, Cambridge, MA, USA}

\begin{abstract}
We analyze the spectral evolution of GRB 090902B and show that subphotospheric
dissipation can explain both the spectra and the spectral evolution.
The emission from a GRB photosphere can give rise to a variety of spectral
shapes. The spectrum can have a shape close to that of a Planck function (as is observed during the first half of GRB090902B) or be broadened, resembling a typical Band function (as is observed during the second half of GRB090902B). The shape mainly depends on the strength and location of the
dissipation in the jet, the ratio of the energy densities of thermal photons and
of the electrons at the dissipation site, as well as on the strength of the
magnetic field.  We further discuss numerical models of the
dissipation and relate these to the observed spectra.  
\end{abstract}

\maketitle

\thispagestyle{fancy}


\section{Introduction}
After more than 40 years of observations the prompt emission in gamma ray bursts is still not fully understood. The most widely accepted model for the emission is the fireball model~\cite{photosphere-models}, \cite{Giann05}, in which the emission is caused by dissipation in a relativistically expanding fireball. This model predicts the presence of a thermal component arising from the photosphere, where the flow which is initially opaque becomes optically thin. 
A reasonable, first assumption is that this component has a shape resembling a Planck function, although this is expected to be slightly modified by relativistic and geometric effects (e.g. \cite{goodman}, \cite{peer08}, \cite{PeerRyde11},  \cite{Beloborodov10},  \cite{Lundman11}). However, the majority of GRB spectra are well fit with a Band function and do not show clear signs of black body emission.  On the other hand,  Planck-like spectra have been observed in a few bursts~\cite{Ryde04, Ryde05} and, in addition, in several cases there is  evidence of a subdominant thermal component accompanying  the non-thermal emission, e.g. in GRB110724B~\cite{Guiriec11}. This has led to a renewed interest in the existence of photospheric emission in GRB spectra.

The presence of dissipation in the GRB outflow is unquestionable, since the spectrum is non-thermal, containing a high energy tail. 
The nature of the dissipation process is, however, not known, but possible scenarios include magnetic reconnection in a Poynting-flux dominated outflow~\cite{Drenkhahn02},\cite{DrenkhahnSpruit02},\cite{Giann05}, internal shocks in the flow~\cite{ReesMeszaros94}, \cite{photosphere-models} and collisional heating~\cite{Beloborodov10}. 
The location of the dissipation is also unknown, but in the internal shock scenario shocks due to colliding shells with different Lorentz factors are expected to occur mainly in the optically thin region of the flow, while oblique shocks can lead to dissipation below the photosphere. In addition, a change in the Lorentz factor of the flow can lead to a shift in the location of the photosphere, so that the dissipation shifts from being located above the photosphere to occurring below the photosphere, with corresponding changes in the observed spectrum. This sub-photospheric dissipation is the subject of the present study.
   
 \begin{figure}[h]
\includegraphics[width=0.45\textwidth]{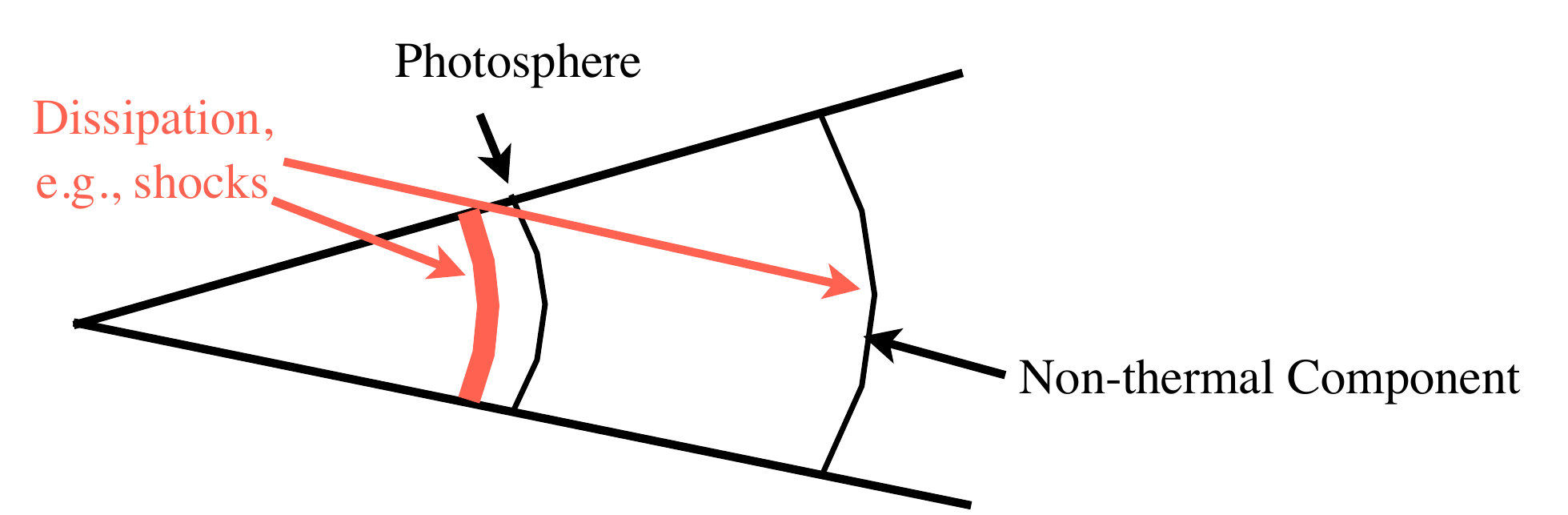}
\caption{Typically the dissipation due to internal shocks occurs above the photosphere, leading to strong non-thermal emission. However, a change in the Lorentz factor of the flow can shift the position of the photosphere, leading to the dissipation site being located below the photosphere.}
\label{Fig:Dissipation}
\end{figure}
 
 \section{Numerical modeling}
 We consider a highly relativistic jet. There is strong thermal emission from the base of the outflow, which is advected outward with the flow and released at the photosphere, where the optical depth drops to unity. Initially the thermal emission takes the form of a Planck function in the comoving frame, and if there is no dissipation below the photosphere the emerging spectrum will also have this shape. Although transformation to the observer frame in addition to geometric effects alters the spectrum somewhat, the observed spectrum should still have a largely Planck-like shape. If, however, part of the kinetic energy is dissipated below the photosphere, the spectrum will be modified and can take on a quite a complex shape~\cite{modelpaper}.
 
We assume that the photosphere is located above the saturation radius where the flow ceases to accelerate, and that the dissipation occurs close to the photosphere, at optical depths of a few.  Whatever the nature of the dissipation process the result is that a population of energetic electrons is created. These cool via synchrotron emission and Compton scattering of low energy photons. The energetic photons which are created through these processes may in turn undergo inverse Compton scattering. In addition pair production and annihilation modifies the particle and photon populations, and affect the other emission processes. The result is a highly non-linear problem, which can only be solved numerically. We compute the resulting spectra using the code described in~\cite{modelpaper}. The kinetic equations are solved self-consistently, taking into account synchrotron emission, SSA, Compton and inverse Compton scattering, pair production/annihilation and electromagnetic cascades. A fraction $\epsilon _d$  of the kinetic energy is assumed to be dissipated -  a fraction $\epsilon _e$ of this goes to acceleration of the electrons, while a fraction $\epsilon _B$ goes to magnetic field generation.

\begin{figure}[h]
\includegraphics[width=0.45\textwidth]{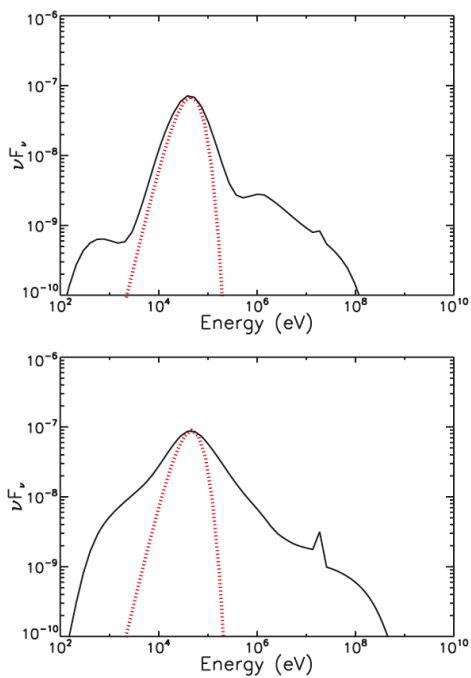}
      \caption{The effect of sub-photospheric dissipation at optical depth $\tau = 10$. The dashed red line represents the initial thermal emission, while the solid black line shows the spectrum which is released at the photosphere in the case of weak (top) and strong (bottom) dissipation. }
      \label{fig:modelspectra}
\end{figure}

Fig.~\ref{fig:modelspectra} show spectra resulting from two dissipation episodes with different characteristics. In both cases the dissipation occurs below the photosphere, at an optical depth of $\tau=10$. The top spectrum results from weak dissipation, with 10 \% of the kinetic energy being dissipation ($\epsilon _d=0.1$). This spectrum has a Planck-like shape. The bottom spectrum is produced by stronger sub-photospheric dissipation ($\epsilon _d=0.2$), which leads to a broader spectrum with a more typical Band-like shape.


\begin{figure}[h]
\includegraphics[width=0.45\textwidth]{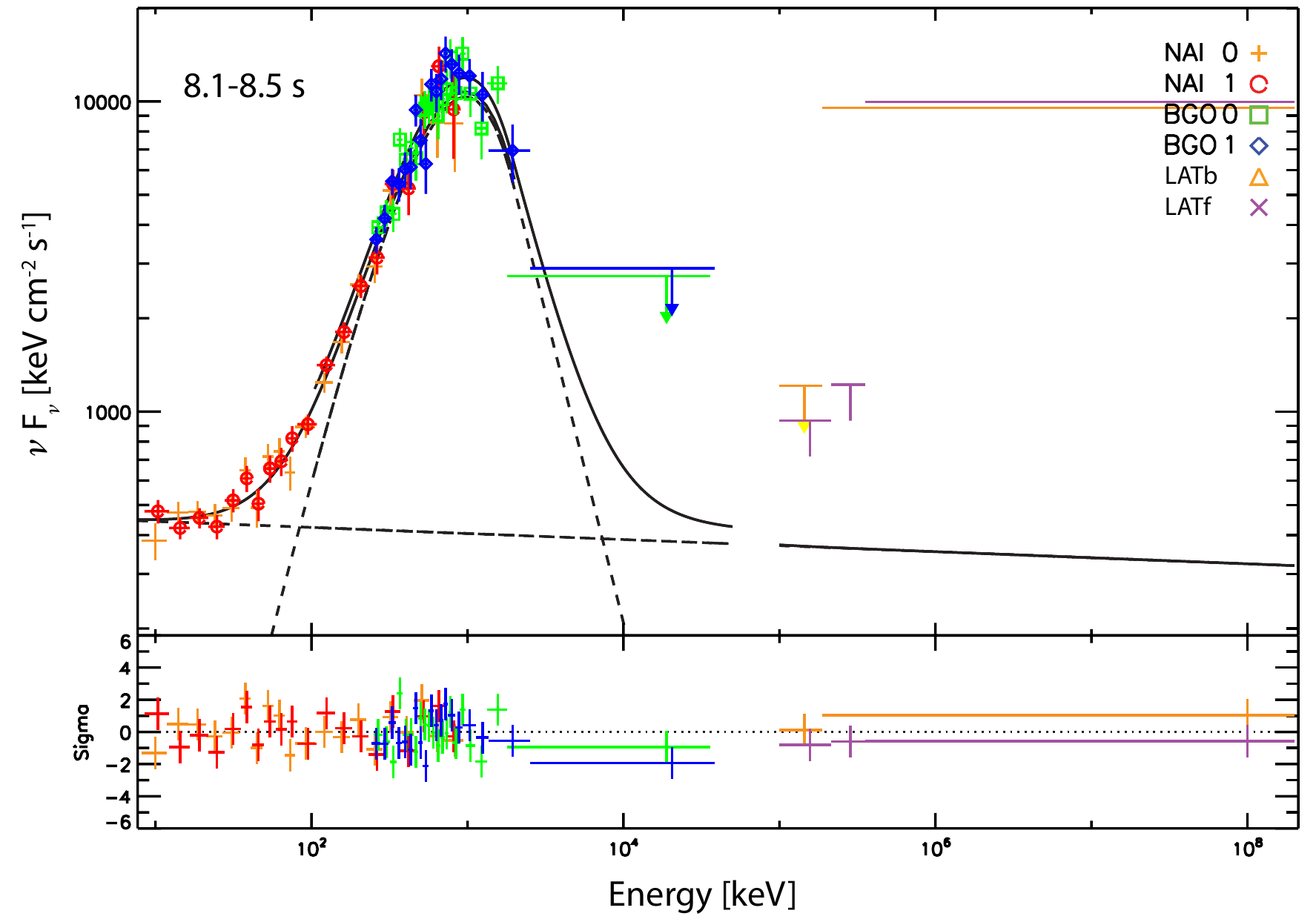}
\includegraphics[width=0.45\textwidth]{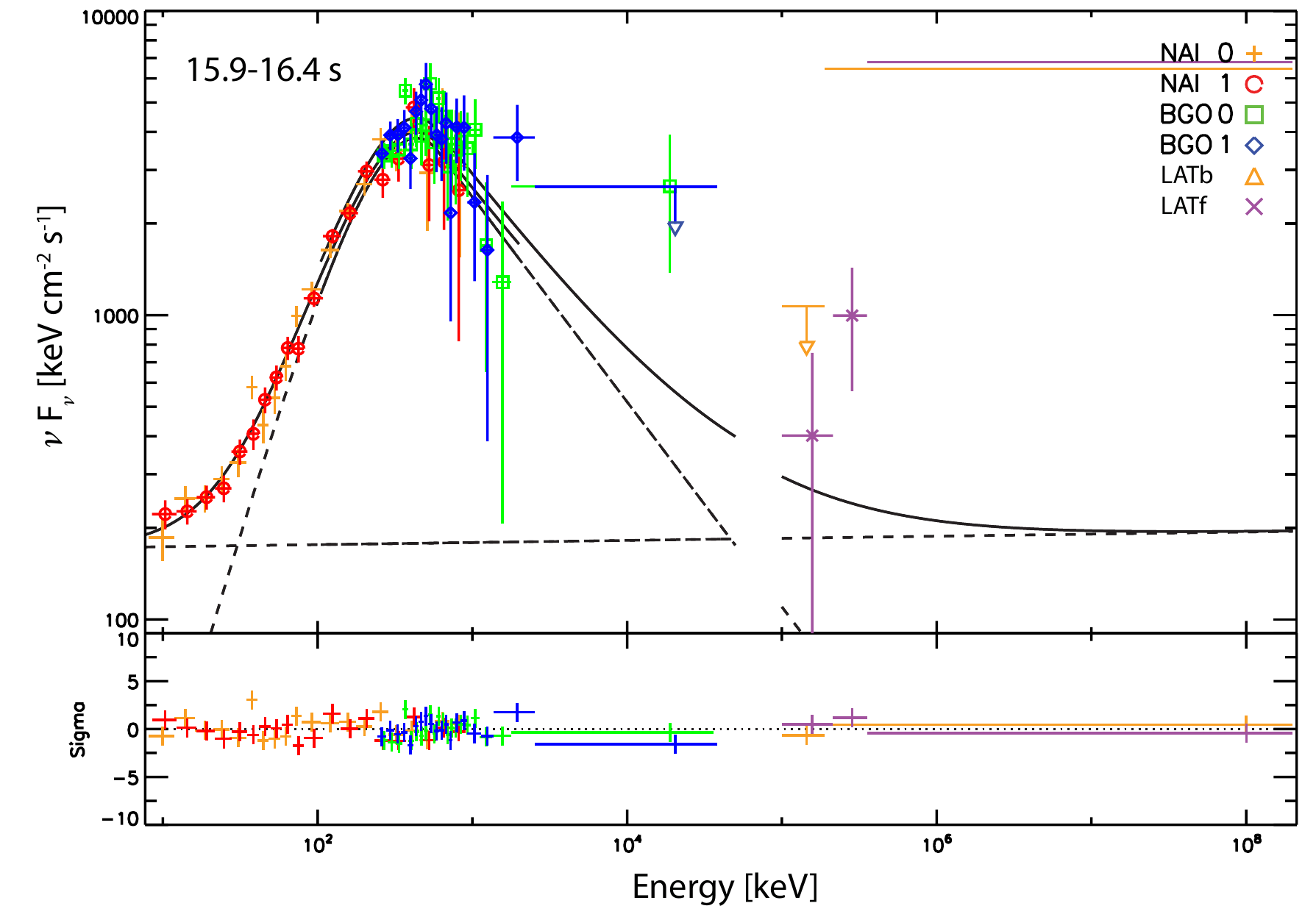}
\caption{Time resolved spectra for two time intervals in epoch 1 and epoch 2 respectively.}
\label{fig:grb090902b-spectra}
\end{figure}

\section{Case study: GRB 090902B}
GRB090902B is one of the strongest gamma ray bursts detected by Fermi.
Emission above 8 keV was detected for approximately 25 s. Throughout the duration of the burst the spectrum has two distinct components; a peaked component at MeV energies and a power law. 
During the first half of the burst the MeV-peak is very steep and narrow (Fig.~\ref{fig:grb090902b-spectra}, top) with a shape close to a Planck function (a better fit is obtained when using a multicolor blackbody, as theoretically expected~\cite{grb090902b-2}). This component is so steep and narrow that it is highly unlikely to have any origin other than the photosphere~\cite{grb090902b-2}. After approximately 12.5~s the spectral characteristics change. The peaked component broadens and is well fit by a typical Band function~\cite{grb090902b-1}, without a significant change in the peak energy or in the flux, resulting in a spectrum which is similar to the spectrum shown in~Fig.~\ref{fig:modelspectra}.  We thus deduce that the most plausible explanation is sub-photospheric dissipation. 

A plausible explanation for the spectral evolution in GRB090902B is therefore that at early times the dissipation occurs above the photosphere, while there is very little dissipation below the photosphere, so that the thermal component escapes largely unmodified. After 12.5 s the dissipation below the photosphere increases in strength, modifying the emerging photospheric component. Such a change could, for instance,  come about through variation in the Lorentz factor of the flow, leading to the  photosphere being located further out, thereby increasing the optical depth at the dissipation site ~\cite{grb090902b-2}.

\section{Conclusion}
Numerical calculations show that the observed spectrum from a GRB photosphere can take on a variety of shapes depending on the strength and location of the dissipation~\cite{PeerMeszarosRees06}. A thermal component 
does therefore not need to be a Planck function, but can in fact mimic a Band function $\le 100$ MeV. 

The case of GRB09092B shows clearly that photospheric emission does occur in gamma ray burst spectra. Moreover, this burst demonstrates that the photosphere can indeed have a variety of shapes. It is thus suggested that a change in dissipation characteristics of this burst  underlies the observed spectral evolution and change the appearance of the spectrum.
We therefore conclude that photospheric emission is probably quite common in GRBs.

\end{document}